# Deep Learning model integrity checking mechanism using watermarking technique

Shahinul Hoque
*University of Tennessee Knoxville*
Knoxville, USA
shoque@vols.utk.edu

Farhin Farhad Riya
*University of Tennessee Knoxville*
Knoxville, USA
friya@vols.utk.edu

Yingyuan Yang
*University of Illinois Springfield*
Springfield, USA
yyang260@uis.edu

Jinyuan Sun
*University of Tennessee Knoxville*
Knoxville, USA
jysun@utk.edu

***Abstract*— In response to the growing popularity of Machine Learning (ML) techniques to solve problems in various industries, various malicious groups have started to target such techniques in their attack plan. However, as ML models are constantly updated with continuous data, it is very hard to monitor the integrity of ML models. One probable solution would be to use hashing techniques. Regardless of how that would mean re-hashing the model each time the model is trained on newer data which is computationally expensive and not a feasible solution for ML models that are trained on continuous data. Therefore, in this paper, we propose a model integrity checking mechanism that uses model watermarking techniques to monitor the integrity of ML models. We then demonstrate that our proposed technique can monitor the integrity of ML models even when the model is further trained on newer data with a low computational cost. Furthermore, the integrity checking mechanism can be used on Deep Learning models that work on complex data distributions such as Cyber-Physical System applications.**

## I. Introduction

In recent years, Machine Learning techniques have become one of the most preferred technique to solve complex real-world problems. More and more systems are integrating Machine Learning and Deep Learning (DL) to solve problems or provide new features. This increase in interest has resulted in more and more organizations to use Machine Learning models. Various Cyber Physical Systems (CPS) now use DL models to improve operation or for various other benefits. Various DL models have been proposed to predict power grid loads, water quality in water treatment plants, predict in and out flow in highway systems, and improving diagnostic tools in healthcare. These systems can be considered as essential and if issues arise in some of these systems, then we might see catastrophic disasters in regional or national level.

Subsequently, to improve operations, CPS systems and other systems are using more and more complex ML techniques and DL models that require large datasets to train and test which is difficult as it is not always possible to obtain large training datasets due to the cost of building such datasets, restrictions, and various privacy concerns. A solution to this difficult situation is Machine Learning as a Service or MLaaS. MLaaS provides a way for organizations having access to large data sources to train ML models and get financial value by providing these DL models to entities that do not have access to large datasets or enough research and development resources. We can already see some cloud computing companies have started to provide dedicated services for power grid operations [11].

CPS architectures deploy such DL models in their essential operation without any way to verify the integrity of the DL model.

Furthermore, we can see that a lot of systems use low computing capable devices as sensors or triggers where such device use pretrained or downloaded DL models. There are also scenarios where large DL models are trained on supercomputers and then are transferred to regular computing capable devices for application. For example, a full build of Autopilot Neural Network in Tesla vehicles involves 48 Networks that take 70,000 GPU hours or around 8 GPU years to train [5]. Then the fully trained Autopilot network is downloaded and deployed by the Tesla vehicles.

We can clearly see a security concerns in such techniques where a system or device is downloading or utilizing a pretrained DL model without checking the integrity of the model. Furthermore, companies are using pre-trained DL models as MLaaS or in a cloud computing environment without monitoring the integrity of the DL model. To solve these problems, we propose a ML model integrity checking mechanism based on watermarking techniques designed for ML models. The integrity checking mechanism can be used to verify the integrity of DL models using a secret key. Our proposed integrity checking mechanism can work even after the DL model has been trained on newer data. This is possible as most of the ML model neural networks are overparameterized and are capable of handling more information even after learning the complex relationship between the input and output data. Our technique uses this overparameterization of Neural Network architectures to store more information inside the DL model and monitor the integrity of the model.

In recent times, we have seen many new proposed watermarking techniques by various research groups for DL models to protect and keep track of intellectual property rights of organizations and ML models. Watermarking techniques provide a solution for ML model owners to present a proof of intellectual property rights on their trained models. In our proposed integrity checking mechanism, we utilize such a watermarking technique by modifying it to work on any DL model deployed for classification problems and make it capable of handling DL models working on any general data distribution.

**Our contributions:** We present a new framework to continuously monitor the integrity of ML models.

We demonstrate that the framework is not dependent on any specific data distribution, and can be adapted to any general ML application

We perform comprehensive experiments to see the changes in performance and accuracy of the model with and without the watermarks embedded.

Finally, we illustrate how our proposed integrity checking mechanism can be integrated in a regular DL model application.

## II. Background

The main purpose of checking the integrity of a file or data is to ensure that a file or data has no unauthorized modifications and that is has not been changed or damaged in any way from the original file or data used in the transportation medium. A common way to ensure the integrity of data files is through hashing which is the process of transforming a data file into a string of fixed length value. The process of hashing a static file is simple and straight forward. However, when we are working with DL models, the models are dynamic in any good application environment the models keep learning and improving with newer data. Is it not feasible to hash a DL model each time it has been trained with newer data.

In recent years we have seen the development of various watermarking techniques for DL models to protect the intellectual property right of the model owners.

Author Uchida et al. [1] proposed to use parameter regularizes to embed watermarks into a DL model in the training phase by imposing certain restrictions on what the model learns. Similarly, Li et al. [2] proposed a watermarking technique that can work in a Black-Box setup environment where the intellectual property of a DL model can be verified by just querying the model using inputs and observing the outputs.

In [4] the authors proposed a watermarking model based on adversarial example and adversarial training. However, an issue with this approach is the ability to generate adversarial examples for the targeted model as it might not always be possible to generate adversarial examples for a given DL model. Similarly, Merrer et al. [6] proposes a technique to embed watermarks in such a way where the watermark can be retrieved using trigger samples without accessing the weights of the Neural Network. [7], [8], [9] propose similar techniques that work in Black-box and White-box setups.

In [12], authors propose a watermarking technique for DL models that use watermarked images to train the DL model to recognize images containing watermarks to predict some targeted labels.

A common issue with the proposed watermarking techniques is that they are targeted for image-based DL models and are dependent on the data distribution type of images as images can have only a fixed value range of 0 to 255 for each color channel. However, as CPS systems are based on a wide range of sensor types of devices that generate values without a fixed range or specific data distribution. Therefore, such techniques are not suitable for CPS applications and their effectiveness in applications designed for CPS systems cannot be guaranteed without further testing and experiments.

## III. Threat Model

### A. Focus

The focus of proposed integrity checking mechanism is to create a mechanism to monitor the integrity of DL models continuously after a fixed time interval. Our proposed approach is capable of verifying the integrity of the model even after training the DL model on new training samples which is not possible when using any hashing technique. Furthermore, our proposed mechanism can be used on any type of DL model for classification-based application.

### B. Assumptions

We assume that an attacker does not have access to the original dataset used to train the model. However, the attacker has access to a shadow dataset that has similar properties to the original dataset used to train the original model. Similarly, we assume the attacker has access to the original DL model architecture, or a DL model architecture very close to the original one. Therefore, our experiments and evaluations have been designed assuming the attacker is capable of training a DL model very close to the original DL model. However, considering the complexity of CPS systems, the immense distribution, and vast categories of sensors, it is unlikely that an attacker can obtain a dataset containing data from all types of sensors, or systems to train such a DL model that an organization trains using their own proprietary dataset.

### C. Non-threats

Our proposed integrity monitoring mechanism does consider all the requirements needed for a watermarking technique as a watermarking technique also needs to provide a way to link the watermark with the model creator's identity to prove the intellectual property rights of the model creator. However, our proposed technique only considers the requirement of checking the integrity of DL models and the resilience of the watermark to verify the integrity of the DL model even after further training the model on newer data.

## IV. System Model

The process of embedding the watermark, generating the Key dataset, and evaluating the integrity of the DL model can be separated in three stages.

The first stage is the regular training stage where the model creator trains a DL model using their own proprietary dataset for a certain application. The second stage is the watermark embedding and Key dataset generation stage. In this, we embed the watermarking by slightly modifying the weights of the DL model and generate the Key dataset based on the modified DL model. This is the most important stage as we need to consider multiple factors based on each individual DL model such as the length of the Key dataset, the embedding epoch. For example, the length of the Key dataset needs to be small compared the length of the original dataset. However, this is not an issue in most circumstances the length of the original dataset is more than 1000 times in larger than the length of the Key dataset.

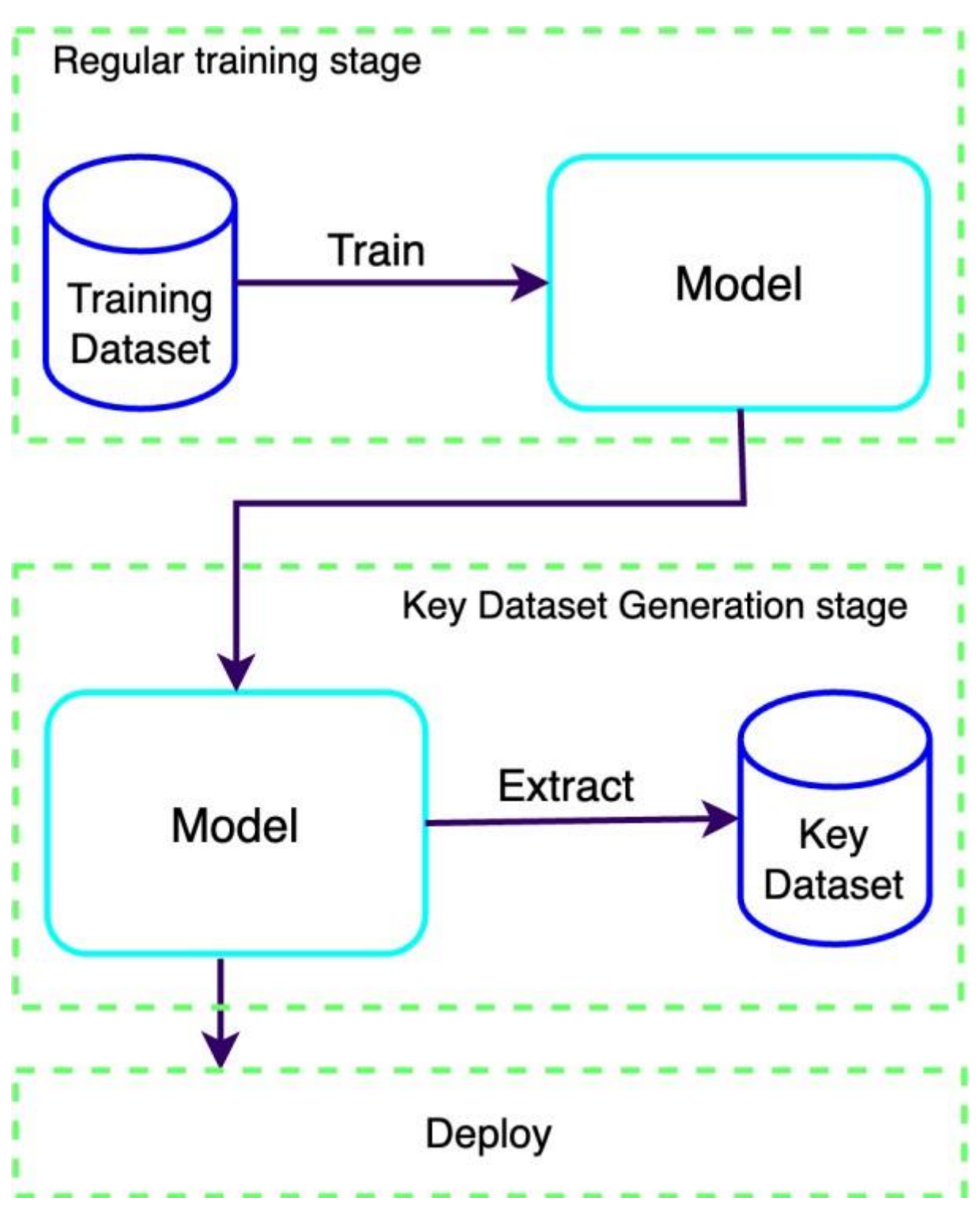

Fig. 1. Diagram of Stage 1 and Stage 2 of the integrity checking mechanism.

How the regular training stage and Key dataset generation stage work are illustrated in Figure 1. In the Key Dataset Generation stage, we generate a dataset consisting of random samples collected from Gaussian distribution computing the length using the Key length $k$. We can consider the length of the random sample dataset (R) as $l$ where $l$ is:

$$l = k * C \quad (□)$$

The constant $C$ remains constant for each specific application and is based on the length of the original dataset (O) used for training the DL model. We also generate a random output $Y^R$ for all the samples in R dataset. Let $Y^{MI}$ be the output of the DL model using the O+R dataset. We select $Y^{MI*}$ using the following criteria:

$$Y^{MI*} =! (Y^R \cap Y^{MI}) \quad (□)$$

Then by training the DL model using the O+R dataset, we observe generate the output $Y^{MA}$. Next, we select all the samples that fall under the following criteria:

$$Y^{MA*} = (Y^R \cap Y^{MA}) \quad (□)$$

We the A and B output, we select all samples that match the following criteria:

$$Y^W = (Y^{MI} \cap Y^{MA}) \quad (□)$$

Then a set of samples along with their label are stored as the Key dataset randomly sampled from the set using the length $k$.

The third stage is the verification stage where an entity can verify the integrity of the model by comparing the predictions of the model using the Key dataset. Because the classes of the key dataset are not based on the original problem data, a DL model that previously did not learn the relationship between the Key data samples and Key labels will not be able to accurately produce the right labels for the Key dataset. Only a DL model trained on the Key Dataset is able to get a high enough accuracy using the Key dataset.

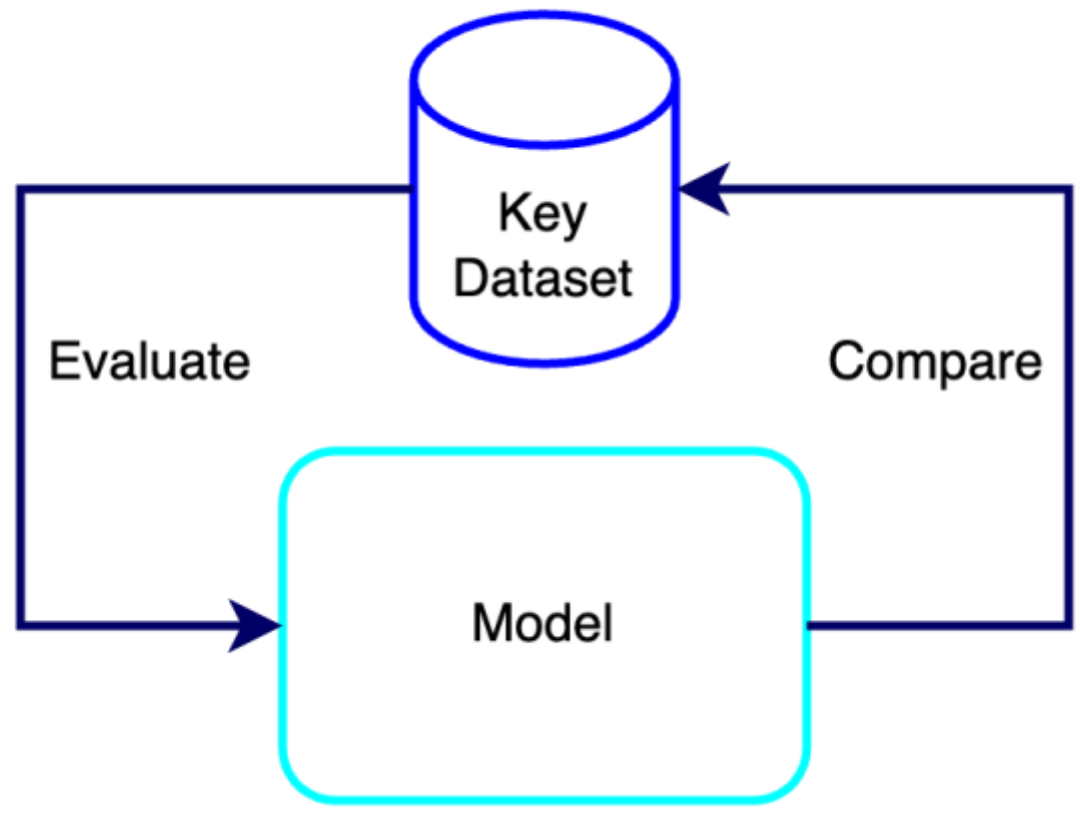

Fig. 2. Diagram of verification stage of the integrity checking mechanism.

We can make our DL model adopt to a new relationship with very low impact on the original problem relationship as all Neural Networks are overparameterized and therefore, they are capable of storing more information than needed.

## V. Implementation

### A. Datasets

In order to analyze the performance of our integrity checking mechanism, we tested our technique on two applications. The first application is a DL model to classify the potability of water in water treatment plans consisting of data records collected from multiple types of sensors recording hardness, conductivity, trihalomethanes, and various other factors contributing to the potability of water [10]. We have selected this application as it is an important part in the water treatment plan which is a critical CPS system. This is an essential system and many people's health directly rely on properly identifying potable water. Furthermore, this system uses data collected from a wide range of sensors and thus difficult to fit using DL models.

How Our second application is a DL model detect anomalies in BUS14 system for Meter reading. The dataset for this application contains Meter readings from various phase angles to detect anomalies.

### B. Experimental Setting

We designed two Neural Network architectures for our chosen two application setups. Our implementation of the DL model is based on the TensorFlow library [13] for Machine Learning. All our experiments were conducted on a M1 Mac device running native TensorFlow library designed for Mac M1 architecture.

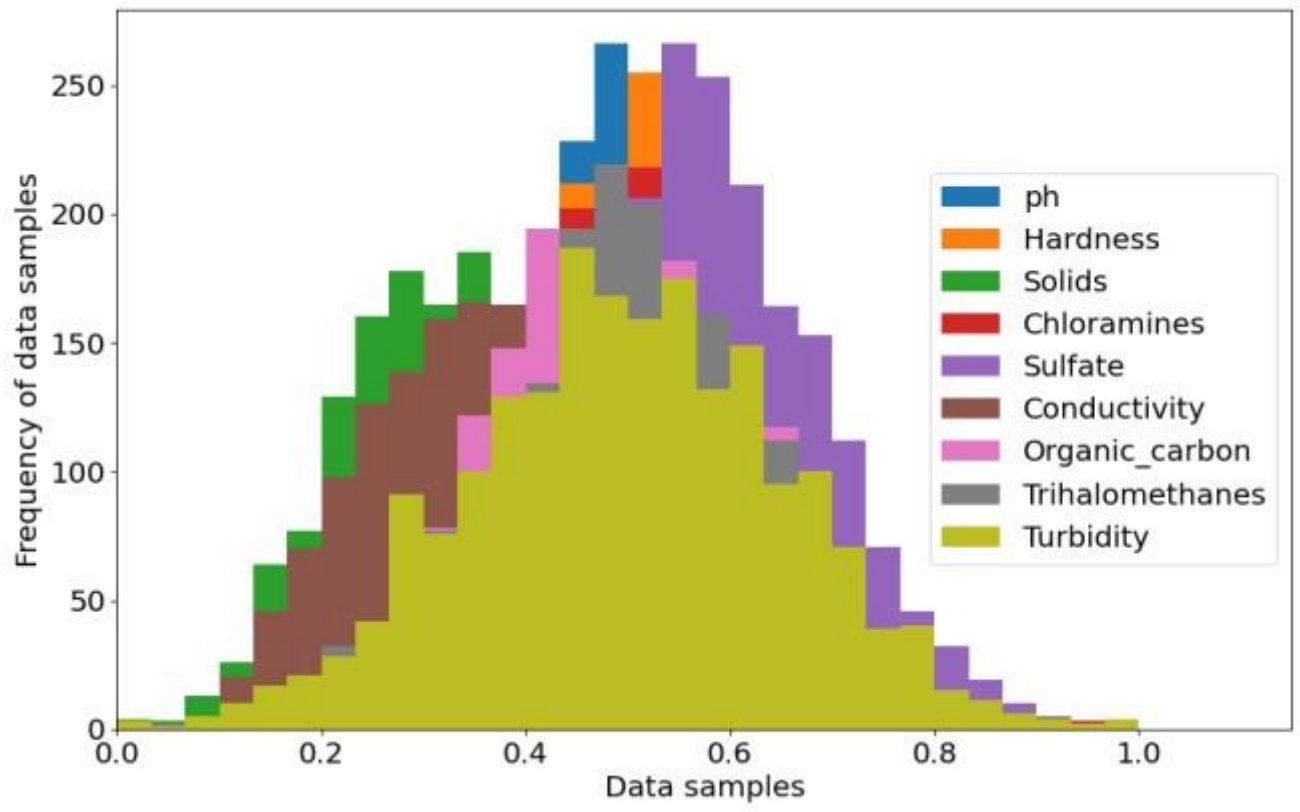

Fig. 3. Histogram of data distribution in Water Quality dataset.

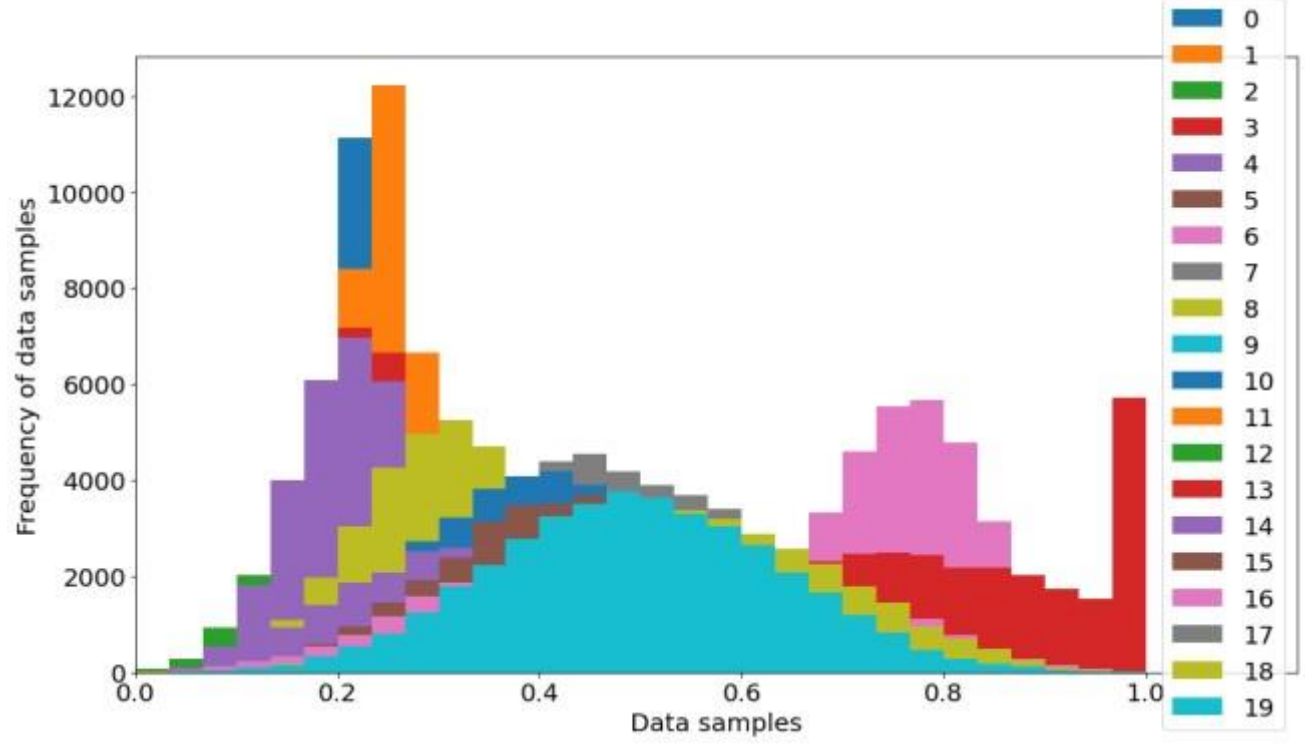

Fig. 4. Histogram of data distribution in BUS14 Anomaly detection dataset.

*C. Data preprocessing*

In order to properly fit our data to our designed Neural Network architecture we decided to normalize the data using the Min-Max normalization technique. Therefore, all the values of both dataset fall under the range of 0 to 1. The process of normalizing the data using either the Min-Max technique or the Mean-Standard Deviation technique is a common practice when preprocessing data. Similarly, in order to generate the Key dataset, we generated all our values between 0 to 1. Therefore, the Key dataset samples and regular dataset samples are indistinguishable, and our technique is not limited to any specific value range such as the proposed techniques that work on image-based applications.

## VI. Experimental Results

In this section we present the experimental results for the DL model integrity checking mechanism.

*A. Water Potability Application*

As we increase the number of embedding epoch to embed the watermarked data more deeply into the DL model, we can see a decrease in non-watermarked model watermark verification accuracy. Therefore, as we more embed the watermark into the DL model, the DL model becomes more distinguishable from a non-watermarked model without much drop in regular application accuracy.

In Table 1, we can see that when we use a longer Key length for our key length dataset, the watermark verification accuracy for a non-watermark model decreases to nearly random detection accuracy. However, increasing the Key length will also result in a drop in the model's regular application accuracy.

TABLE I. Effect of Key-Length in Watermark detection accuracy

| Key length | Model Accuracy | Watermark detection accuracy for non-watermarked model |
|---|---|---|
| 10 | 67.12% | 83.99% |
| 20 | 68.71% | 78.01% |
| 30 | 67.32% | 69.99% |
| 40 | 65.84% | 71.49% |
| 50 | 65.34% | 65.99% |
| 60 | 63.36% | 62.66% |
| 70 | 665.24% | 63.01% |
| 80 | 663.16% | 64.24% |
| 90 | 61.97% | 62.44% |
| 100 | 61.06% | 63.01% |

## VII. Conclusion

Nowadays, we can see an increase in the deployment of pre-trained DL models. Many critical systems such as Cyber-Physical systems related to the power grid, water treatment plant and many other have started to integrate such pretrained model in their day-to-day applications. As such, our proposed integrity checking mechanism provides a way for verifying the integrity of the pre-trained model after deployment.

Now that many essential services are integrated using DL models, monitoring the integrity of the DL model could help preventing blackouts from such services either due to attacks or integrity degradation of DL models.

Moreover, our proposed technique can be generalized to be able to work in a classical Machine Learning application environment. Also, the watermarking scheme used by our technique is not reliable on any specific data distribution and therefore, is not limited to DL models utilized in only specific applications such as image processing, or natural language processing.

Furthermore, an interesting direction for the future would be to replace the last layer of prediction-based DL models using static multi-neuron layers, so that we can implement our integrity checking mechanism to prediction-based DL models.